# To the theory of interaction between electron and nuclear systems.

V.L.Marchenko, A.M.Savchenko

Coupled electron-nuclear oscillations in antiferromagnetics with anisotropy of "light plane" type in the strong external magnetic field are under consideration. The new mode $\varepsilon_{2k}$ of these coupled oscillations is obtained for antiferromagnetic systems by using «u-v»-Bogolubov's unitary transformations. The dynamic shift of the frequency of nuclear magnetic resonance concerned with this mode is obtained.

PACS:74.70.H

Keywords: Antiferromagnetic, Magnetic field, Oscillations, Magnetostriction

It's well known that an increase of the constant external magnetic field leads to the gradual collapse of spins of magnetic sublattices in antiferromagnetic. The second order transition takes place in the strong enough magnetic field: magnetic moments line up along the field (paramagnetic phase). Moreover, in the vicinity of the critical collapse field $H_c$ long-wave collective oscillations in the nuclear spin system are in existence. These are nuclear spin waves initiated by electron spin waves of the quasiantiferromagnetic branch. The aim of this work is to investigate the interaction between nuclear spin oscillations and electron spin waves.

Let's consider the Hamiltonian of our model in the form [1,2]:

$$H=\sum_{f,g}\left[J\left(\vec r_{fg}\right)\vec S_f\vec S_g+K_2\left(\vec S_f\vec n\right)\left(\vec S_g\vec n\right)\right]+K_1\left[\sum_f\left(\vec S_f\vec n\right)^2\sum_g\left(\vec S_g\vec n\right)^2\right]+\mu\vec H\left(\sum_f\vec S_f+\sum_g\vec S_g\right)-$$
$$-\mu_n\vec H\left(\sum_f\vec I_f+\sum_g\vec I_g\right)-A_0\left(\sum_f\vec I_f\vec S_f+\sum_g\vec I_g\vec S_g\right) \qquad (1)$$

where electron spins $\vec S_f,\vec S_g$ belong to different sublattices "f" and "g", $\vec I_f,\vec I_g$ - nuclear spins, $J\left(\vec r_{fg}\right)$ - the integral of exchange interaction, $\vec r_{fg}=\vec r_f-\vec r_g$, $K_1,K_2$ - anisotropy constants, $\vec n$ - normal to plane of light magnetization, $A_0$- the constant of electron-nuclear interaction.

Now represent our Hamiltonian using the second quantization operators. We can express operators [3] $\vec S_j$ using operators of spin deflections $\mathrm{a}_j^+,\mathrm{a}_j$ with the help of Holstein-Primakoff transformation. Further apply Fourier representation

$$\mathrm{a}_{\nu\vec k}=N^{-1/2}\sum \mathrm{a}_j e^{i\vec k\vec r_j},\nu=\begin{cases}1, j=f\\2, j=g\end{cases}$$

and introduce new operators $\tilde{\mathrm{a}}_{\nu\vec k}=2^{-1/2}[\mathrm{a}_{1\vec k}+(-1)^{\nu-1}\mathrm{a}_{2\vec k}]$. So the Hamiltonian of our model assume the form

$$H=\sum_{\vec k,\nu}\left\{A_{\nu\vec k}\tilde{\mathrm{a}}^+_{\nu\vec k}\tilde{\mathrm{a}}_{\nu\vec k}+\frac{1}{2}B_{\nu\vec k}(\tilde{\mathrm{a}}_{\nu\vec k}\tilde{\mathrm{a}}_{\nu-\vec k}+\tilde{\mathrm{a}}^+_{\nu\vec k}\tilde{\mathrm{a}}^+_{\nu-\vec k})\right\}-\omega_n(\sum_f I_f^{z''}+\sum_g I_g^{z''''}+\frac{1}{2}A_0S^{1/2}\sum_{\vec k,\nu}\left\{\tilde{\mathrm{a}}_{\nu\vec k}[X_1+(-1)^{\nu-1}X_2]+э.с.\right\}$$

where

$$A_{1\vec k}+A_{2\vec k}=2J(0)S\cos2\theta+2K_1S+2A_0\langle I\rangle\cos(\theta-\theta_n)+2\mu H\sin\theta$$

$$A_{1\vec k}-A_{2\vec k}=2J(\vec k)S\sin^2\theta-K_2S$$

$$B_{1\vec k}+B_{2\vec k}=-2K_1S$$

$$B_{1\vec{k}} - B_{2\vec{k}} = -2J(\vec{k})S\cos^2\theta + K_2 S, J(\vec{k}) = \sum_f J(r_{fg})e^{i\vec{k}\vec{r}_{fg}}$$

$$X_1 = N^{-1/2}\sum_f e^{i\vec{k}\vec{r}_f}[(I_f^{z'n} - \langle I\rangle)\sin(\theta-\theta_n) - I_f^{x'n}\cos(\theta-\theta_n) + iI_f^{y'n}]$$

$$X_2 = N^{-1/2}\sum_g e^{i\vec{k}\vec{r}_g}[-(I_g^{z''n} - \langle I\rangle)\sin(\theta-\theta_n) - I_g^{x''n}\cos(\theta-\theta_n) + iI_g^{y''n}]$$

Electron and nuclear spin systems of double-sublattice antiferromagnetic have four resonance frequencies which correspond to similar types of spin precession. With neglect of dynamic mode coupling of these systems nuclear resonance frequencies prove to be degenerate and are equal to $\omega_n$. If we take into account this dynamic coupling we obtain the shift of resonance frequencies, which is for nuclear frequency inversely proportional to electron resonance frequency squared.

Activation energy of spin waves of quasiantiferromagnetic branch in antiferromagnetics of the "light plane" type [4] (in magnetic fields $H \ll H_E$) is caused by magnetic anisotropy and crystal exchange interaction field, $\varepsilon_{20} \cong \mu(2H_A H_E)^{1/2}$. While the activation energy of spin waves of quasiferromagnetic branch in these magnetic fields is already caused by spontaneous magnetostriction, exchange energy and hyperfine interaction, $\varepsilon_{10} \cong \mu[2H_E(H_{ms} + H_{N0})]^{1/2}$. It follows that the shift of the nuclear resonance frequency coupled with the quasiantiferromagnetic mode in magnetic fields $H \ll H_E$ is sufficiently great (about ten percents).

$\varepsilon_{20} \ll \varepsilon_{10}$, if $H$ and $H_c$ are of the same order, therefore, in this case the dynamic shift of nuclear magnetic resonance frequency coupled with quasiantiferromagnetic mode is quite possible.

And exactly in this case long-wave collective oscillations (in the nuclear spin system), so called nuclear spin waves, exist. They interact with electron spin waves of the quasiantiferromagnetic branch and hereinafter just this phenomenon is under consideration.

Let's determine the spectrum of coupled oscillations of electron and spin systems. Apply Holstein-Primakoff transformation to nuclear spin operators $\vec{I}_j, j=f,g,$ using Bose-operators $\alpha_j^+,\alpha_j$. Keeping in the Hamiltonian only terms quadratic in nuclear spin operators, we obtain

$$H=\sum_{\vec{k}}\left\{A_{2\vec{k}}\tilde{a}^+_{1\vec{k}}a_{1\vec{k}}+\frac{1}{2}B_{2\vec{k}}(\tilde{a}_{2\vec{k}}\tilde{a}_{2-\vec{k}}+\tilde{a}^+_{2\vec{k}}\tilde{a}^+_{2-\vec{k}})+\omega_n\alpha^+_{2\vec{k}}\alpha_{2\vec{k}}\right\}+\sum_{\vec{k}}\left\{C_{2\vec{k}}\tilde{a}^+_{1\vec{k}}\alpha^+_{2\vec{k}}+D_{2\vec{k}}\tilde{a}_{2\vec{k}}\alpha_{2-\vec{k}}+\omega_n\alpha_{2\vec{k}}+\text{э.с.}\right\} \quad (2)$$

где

$$\alpha_{2\vec{k}}=2^{-1/2}(\tilde{\alpha}_{1\vec{k}}-\tilde{\alpha}_{2\vec{k}}),C_{2\vec{k}}=\frac{1}{2}[1-\cos(\theta-\theta_n)](\omega_{n0}\omega_N)^{1/2},$$

$$D_{2\vec{k}}=-\frac{1}{2}[1+\cos(\theta-\theta_n)](\omega_{n0}\omega_N)^{1/2}$$

We can diagonalize the quadratic form (2) by using «u-v» Bogolubov's unitary transformation [5]

$$\tilde{a}_{2\vec{k}}=u_{ee\vec{k}}C_{2e\vec{k}}+v_{ee\vec{k}}C^+_{2e-\vec{k}}+u_{en\vec{k}}C_{2n\vec{k}}+v_{en\vec{k}}C^+_{2n-\vec{k}},$$

$$\alpha_{2\vec{k}}=u_{ne\vec{k}}C_{2e\vec{k}}+v_{ne\vec{k}}C^+_{2e-\vec{k}}+u_{nn\vec{k}}C_{2n\vec{k}}+v_{nn\vec{k}}C^+_{2n-\vec{k}},$$

where functions u и v are defined by

$$u_{ne\vec{k}}=\left(\frac{\omega_{N0}}{\Omega_{2n\vec{k}}}\right)^{1/2}\frac{\omega_n A_{2\vec{k}}}{\varepsilon^2_{2\vec{k}}},\ v_{en\vec{k}}=\left(\frac{\omega_{N0}}{\Omega_{2n\vec{k}}}\right)^{1/2}\frac{\omega_n B_{2\vec{k}}}{\varepsilon^2_{2\vec{k}}},$$

$$u_{nn\vec{k}}=\frac{(\omega_n\omega_{N0})^{1/2}}{\omega_n-\Omega_{2n\vec{k}}}u_{ne\vec{k}},\ v_{nn\vec{k}}=\frac{(\omega_n\omega_{N0})^{1/2}}{\omega_n+\Omega_{2n\vec{k}}}v_{en\vec{k}},$$

$$u_{ee\vec{k}}=\left(\frac{A_{2\vec{k}}+\varepsilon_{2\vec{k}}}{2\varepsilon_{2\vec{k}}}\right)^{1/2},\ v_{ee\vec{k}}=-\left(\frac{A_{2\vec{k}}-\varepsilon_{2\vec{k}}}{2\varepsilon_{2\vec{k}}}\right)^{1/2},$$

$$u_{ne\vec{k}}=\frac{(\omega_n\omega_{N0})^{1/2}}{\omega_n-\Omega_{2e\vec{k}}}u_{ee\vec{k}},\ v_{ne\vec{k}}=\frac{(\omega_n\omega_{N0})^{1/2}}{\omega_n+\Omega_{2e\vec{k}}}v_{ee\vec{k}}.$$

And so our Hamiltonian assumes the form

$$\tilde{H}=\sum_{\vec{k}}(\Omega_{2e\vec{k}}C^+_{2e\vec{k}}C_{2e\vec{k}}+\Omega_{2n\vec{k}}C^+_{2n\vec{k}}C_{2n\vec{k}}),$$

where $C^+_{2e\vec{k}},C_{2e\vec{k}},C^+_{2n\vec{k}},C_{2n\vec{k}}$ - creation and destruction operators of normal modes of quasielectron and quasinuclear spin waves with frequencies $\Omega_{2e\vec{k}}$ и $\Omega_{2n\vec{k}}$, which can be obtained from the dispersion equation

$$\left(\Omega^2-\varepsilon_{2\vec{k}}^2\right)\left(\Omega^2-\omega_n^2\right)+2\Omega^2\left(C_{2\vec{k}}^2-D_{2\vec{k}}^2\right)-2A_{2\vec{k}}\omega_n(C_{2\vec{k}}^2+D_{2\vec{k}}^2)+$$
$$+4B_{2\vec{k}}C_{2\vec{k}}D_{2\vec{k}}\omega_n+\left(C_{2\vec{k}}^2-D_{2\vec{k}}^2\right)^2=0$$

If we take into account the static influence of the nuclear subsystem we'll obtain the energy gap in electron spectrum $\varepsilon_{2k}$ (for $k=0$):

$$\varepsilon_{2\vec{k}}=\left(A_{2\vec{k}}^2-B_{2\vec{k}}^2\right)^{1/2}=\mu\left\{H\sin\theta+H_N\cos\left(\theta-\theta_n\right)+2J(0)S\mu^{-1}\cos2\theta-\mu^{-1}[J(\vec{k})-J(0)S\right\}^{1/2}\times$$
$$\times\left\{H\sin\theta+H_A+H_N\cos(\theta-\theta_n)+2\mu^{-1}J(0)S-\mu^{-1}[J(\vec{k})-J(0)]S\right\}^{1/2}$$

The result for $\Omega_{2k}$ in the case of antiferromagnetic of "light plane" type in a long-wave approximation for magnetic fields $H \ll H_c$ can be written in the form

$$\Omega_{2\vec{k}}^2=\frac{1}{2}\left\{\omega_n^2+\varepsilon_{2\vec{k}}^2+(-1)^j2\omega_N\omega_{n0}\pm\left[(\omega_n^2+\varepsilon_{2\vec{k}}^2)^2+4(-1)^j\omega_N\omega_{no}(\omega_n^2+\varepsilon_{2\vec{k}}^2)+8\omega_n\omega_{n0}\omega_NA_{2\vec{k}}\right]^{1/2}\right\}$$

where $j=1,2$.

It follows from the last formula that the dynamic coupling of electron and nuclear systems leads to the nullification of one of two frequencies of normal oscillations in the point of phase transformation to paramagnetic state. In this phase one of spin-wave branches is high activative and the other - low activative in the vicinity of phase transition point to paramagnetic state.

If $\varepsilon_{20}^2 \gg 4\max\left\{\omega_n^2,\omega_n\omega_{N0}\right\}$, the spectrum of quasielectron (quasinuclear) oscillations for the "light plane" type antiferromagnetic can be presented in the form

$$\Omega_{2\vec{k}} \cong \begin{Bmatrix} \varepsilon_{2\vec{k}} \\ \omega_n \left(1 - \frac{2A_{2\vec{k}}\omega_{N0}}{\varepsilon_{2\vec{k}}^2}\right)^{1/2} \end{Bmatrix}$$

In this case there are two oscillation branches. One –electron and one – quasinuclear.

## References.